\documentclass[10pt,conference]{IEEEtran} 
\IEEEoverridecommandlockouts
\usepackage{cite}
\usepackage{amsmath,amssymb,amsfonts}
\usepackage{hyperref}
\usepackage{algorithmic}
\usepackage{graphicx}
\usepackage{textcomp}
\usepackage{xcolor}

\usepackage{siunitx}

\usepackage{braket}
\def\BibTeX{{\rm B\kern-.05em{\sc i\kern-.025em b}\kern-.08em
    T\kern-.1667em\lower.7ex\hbox{E}\kern-.125emX}}

\begin{document}

\title{One nine availability of a Photonic Quantum Computer on the Cloud toward HPC integration}

\author{\IEEEauthorblockN{
Nicolas Maring\IEEEauthorrefmark{1},
Andreas Fyrillas\IEEEauthorrefmark{1}\IEEEauthorrefmark{2}, 
Mathias Pont\IEEEauthorrefmark{1}, 
Edouard Ivanov\IEEEauthorrefmark{1},\\
Eric Bertasi\IEEEauthorrefmark{1},
Mario Valdivia\IEEEauthorrefmark{1} and
Jean Senellart\IEEEauthorrefmark{1}
}
\IEEEauthorblockA{\IEEEauthorrefmark{1}Quandela, Massy, France\\
Email: jean.senellart@quandela.com} \IEEEauthorblockA{\IEEEauthorrefmark{2}Centre for Nanosciences and Nanotechnologies, CNRS, Universit\'e Paris-Saclay, Palaiseau, France}
}

\maketitle

\begin{abstract}
The integration of Quantum Computers (QC) within High-Performance Computing (HPC) environments holds significant promise for solving real-world problems by leveraging the strengths of both computational paradigms. However, the integration of a complex QC platform in an HPC infrastructure poses several challenges, such as operation stability in non-laboratory like environments, and scarce access for maintenance. Currently, only a few fully-assembled QCs currently exist worldwide, employing highly heterogeneous and cutting-edge technologies. These platforms are mostly used for research purposes, and often bear closer resemblance to laboratory assemblies rather than production-ready, stable, and consistently-performing turnkey machines. Moreover, public cloud services with access to such quantum computers are scarce and their availability is generally limited to few days per week. In November 2022, we introduced the first cloud-accessible general-purpose quantum computer based on single photons. One of the key objectives was to maintain the platform's availability as high as possible while anticipating seamless compatibility with HPC hosting environment. In this article, we describe the design and implementation of our cloud-accessible quantum computing platform, and demonstrate one nine availability (92\%) for external users during a six-month period, higher than most online services. This work lay the foundation for advancing quantum computing accessibility and usability in hybrid HPC-QC infrastructures.

\end{abstract}


\section{Introduction}

The first quantum computer available on the cloud was introduced by IBM in May 2016 \cite{ibm2016}. IBM's quantum computing platform, known as IBM Quantum Experience, provided researchers and developers with remote access to a small-scale quantum computer through the cloud. This marked an important milestone in the democratization of quantum computing, allowing users to experiment and explore quantum algorithms without the need for physical access to the hardware \cite{devitt2016performing}. Since then, several other companies and organizations have also made quantum computers accessible through direct cloud-based platforms or third-party services \cite{hassija2020present}, contributing to the growth and accessibility of quantum computing.  

However, these resources have relatively low availability, with system uptime\footnote{as evident from the device availability dashboards of different platforms, for instance AWS Console {\tt AWS Braket > Device} (\url{https://us-east-1.console.aws.amazon.com/braket/home\#/devices}), or Microsoft Azure {\tt AzureQuantum > Providers} } generally ranging from one day to four days per week, which falls far below the high availability standards set for classical resources on the same platforms (e.g., three to five nines). For all the providers, the reasons for this limited availability are likely rooted in multiple challenges:

\begin{itemize}
    \item Limited hardware resources: quantum computers are in the early stages of their development, and the number of available platforms is limited. This scarcity leads to a high demand for access to quantum computers, resulting in limited availability for users.
    \item Complex hardware maintenance: quantum computers require specialized maintenance and calibration procedures to ensure their proper functioning. These processes are time-consuming and require dedicated, highly trained teams.
    \item Error rates and stability: quantum computers are highly sensitive to environmental noise and susceptible to errors. Achieving high stability and low error rates over a long period of time is a significant challenge in quantum computing. When errors occur, it requires troubleshooting and recalibration, resulting in periods of unavailability.
    \item Research and development focus: many quantum computers available through cloud services are still in the research and development phase (TRL~4 or above). These systems may undergo frequent updates, modifications, and experimentation, leading to intermittent periods of unavailability as new features or improvements are implemented.
\end{itemize}

We released in November 2022 the first single-photon based computing platform named Ascella \cite{maring2023general}, and provided free online access through the Quandela Cloud service \cite{QuandelaCloud}. In this work, we present the architecture of Ascella and the specific challenges to making the platform openly accessible. Inspired from the ``early days'' of classical computers \cite{gray1991high}, we describe the protocols, processes, and adaptations implemented to achieve a \textit{first nine} availability of the service. Furthermore, we provide an overview of the remaining challenges expected for scaling up to larger systems and a roadmap towards achieving high-performance computing (HPC) standards. Lastly, we emphasize the necessity of transparency in availability and real-time performance metrics for the possible and efficient integration of QC into HPC environment.

\section{Architecture and key operation challenges}
\label{sec:architecture}
\begin{figure*}[]
    \centering
    \includegraphics[width=0.90\linewidth]{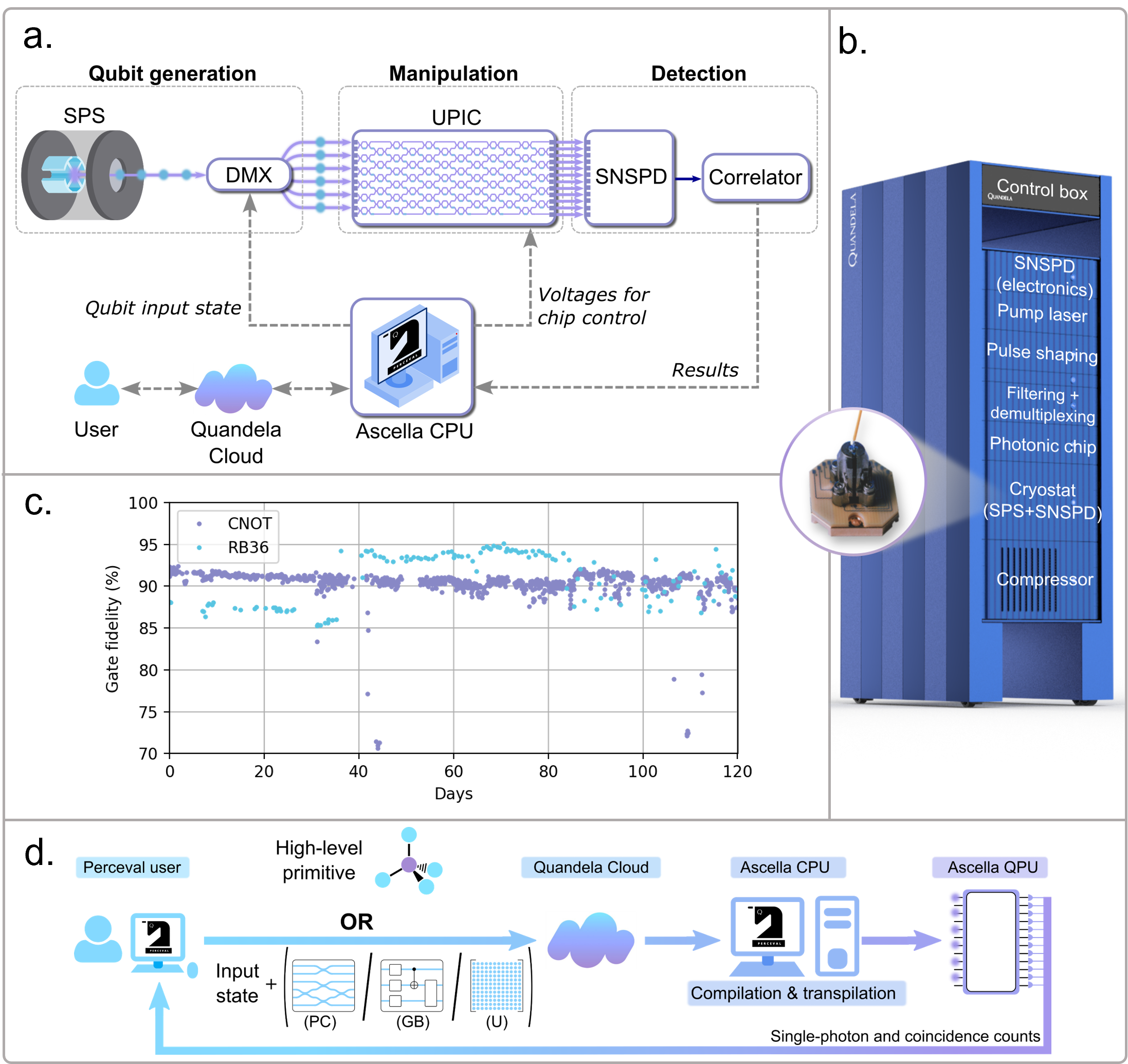}
    \caption{
    \textbf{Architecture of the \textit{Ascella} QPU}
    \textbf{a.} A  quantum-dot single-photon source (SPS) device at \qty{5}{\kelvin} is operated at \qty{80}{\mega\hertz} repetition rate. An active optical demultiplexer (DMX) followed by fibered delays converts the train of single photons into 6 photons arriving simultaneously to the 12-mode universal photonic integrated chip (UPIC). Photons are detected at the chip output by superconducting nanowire single-photon detectors (SNSPD) and detection times are processed by a correlator.
    A full software stack controls the unitary matrix implemented on the chip through the voltages applied on 126 thermo-optic phase shifters and the photonic input state according to the job requested. 
    \textbf{b.} The setup may be integrated into a server-compatible structure.
    \textbf{c.} Continuous benchmark over 120 days of operation. CNOT is the average of a simple CNOT fidelity run at different position on the photonic chip, and RB36 operates a full tomography of a CNOT gate through 36 sampling taks as detailed in \cite{maring2023general} "\emph{Benchmarking logic gates}". 
    \textbf{d.} \textit{Perceval} users may send jobs to the Quandela Cloud consisting in either a high-level primitive, like a variational quantum eigensolver for a molecule, or an input state along with a photonic circuits (PC), a gate-based circuit (GB), or a unitary matrix (U). The job is first processed by a CPU, which computes the optimized voltages for the on-chip phase shifters from our compilation and transpilation process. Finally, the user receives the collected single-photon and coincidence counts after the computation on the quantum processing unit (QPU).
    }
    \label{fig:main_hardware}
\end{figure*}

Ascella's hardware, as shown in Fig.\,\ref{fig:main_hardware}.a, is based on an on-demand high-brightness source of pure and indistinguishable single photons~\cite{somaschi2016near}. This source, made from a semiconductor quantum dot (behaving as ``an artificial atom'') embedded in a micropillar cavity, requires both an applied bias voltage and a finely tuned excitation laser to emit pure single photons \cite{Thomas2021c}. These are then interfaced with a deterministic optical demultiplexer~\cite{pont2022quantifying}, generating and initializing up to 6 path-encoded (dual-rail) qubits. Photon arrival times are synchronized using fiber delay lines, ensuring on-chip photons wave packets overlap. The qubits are manipulated in a 12-mode fully re-configurable universal photonic integrated circuit (UPIC)~\cite{taballione202012}, and measured with high efficiency single-photon detectors and a post-processing unit.

To operate the machine, tasks are sent by the user to Quandela Cloud through standard REST API using the python-based open-source framework 
\textit{Perceval}~\cite{heurtel2023perceval}. The user can either specify an available high-level primitive (for instance photonic Variational Quantum Eigensolver on a specific molecule), a photonic circuit, a gate-based circuit or a unitary transformation to be applied to a specified input state containing $1$ to $6$ photonic qubits, and optional post-selection criteria. Output photon coincidences are then acquired up to the desired sample number and data sample results are sent back to the user to be interpreted in the context of the running algorithm.

As common in a research facility, the full set-up was designed and assembled on an optical table, as a mosaic of in-house and off-the-shelf components (laser, cryostats, mirrors, lenses, optical fibers, etc...). Active components are connected to a single desktop computer through USB and controlled through heterogeneous proprietary software stacks and ad-hoc communication protocols. Most of the components are operated at room temperature with standard air conditioning. To enable a stable operation of the cloud-available system, it is necessary to consider hardware faults and fluctuating performance.
Indeed, the quantum computing system consists of various components susceptible to faults such as communication errors, feed-back control instability, or internal software bugs. The main challenge is that these failures may not break the entire system flow, leading to the production of inconsistent results which the end-user may not even be aware of. Anticipating hardware faults is essential to ensuring the stability and reliability of the system.
Furthermore, hardware components involved in generating and manipulating qubits are sensitive to environmental fluctuations. For instance, spikes in current can modify the voltage bias applied to the single-photon source, typical room temperature fluctuations can alter the refractive index of delay fibers, causing slight delays between photons and polarization rotations, and room humidity fluctuations can impact free-space modules. Managing these fluctuations and their impact on the system's performance is crucial for maintaining consistent and reliable results.

The seamless user experience of our cloud-accessible computer leads to additional constraints. For a cloud-based service, users outside our close collaborators must be taken into account: it is imperative to maintain consistent availability and responsiveness, even during non-working hours, as well as programmed and openly accessible system updates and maintenance windows. Ensuring seamless communication, setting realistic expectations, and providing prompt support is vital to user satisfaction.

\section{Processes}

To enhance hardware reliability, expedite maintenance response time, and elevate user experience, we implemented a series of processes.

\subsection{Monitoring and alerts}
\label{sec:monitoring}
Every hour, Ascella undergoes a few minutes automated calibration where 92 system parameters and metrics, such as optical transmissions, single-photon purity and indistinguishability~\cite{ollivier2021hong}, and modules temperatures, are measured. Note that this calibration period is counted as down-time in our availability estimation. The metrics are sent to an aggregation server called \textit{Minotaur}, which includes a MQTT server \textit{Mosquitto} and the standardized data aggregation server \textit{Prometheus}. The logs are exported to a log data platform\footnote{\url{https://www.ovhcloud.com/en/logs-data-platform/}} utilizing the \textit{graylog} standard. All metrics are visualized on a dashboard using a \textit{Grafana} visualization server and have been displayed in break rooms and high-visibility locations. Based on predetermined thresholds on key metrics, the Grafana alert engine sends anomaly notifications to the cloud operations team.
This tool plays a crucial role in rapidly diagnosing failures and facilitating swift maintenance actions. 

Finally, we collect ``usage data'': all the technical information related to user tasks sent to the platform (in average \num{32000} per month in the first 6 months). Usage data is stored preserving only timestamp, time of operation, compiled circuit, input photon configuration and task numeric results. All these logs are kept with a 6 month retention period and used for system stability and predictive maintenance.


\subsection{Methodological reaction and feedback control system}
\label{sec:feedback}
Our approach to mitigating the fluctuating performance discussed in Sec.~\ref{sec:architecture} -- and minimizing the frequency of raised alerts -- is based on an agile-like incident management methodology. This methodology relies on incident triage and tracking, down-time transparency, swift reaction, root cause analysis, monitoring (see Sec.~\ref{sec:monitoring}), automated testing and continuous improvement. In particular, we use active feedback loops on a number of parameters in the opto-electronic system. These loops are kept as simple as possible to ensure robustness against measurement noise: they mostly consist of coordinate descents with fixed-step line searches. To proactively anticipate performance decline and expedite corrective measures, we have identified a set of five pivotal parameters in the qubit generation and manipulation modules (see Fig.\ \ref{fig:main_hardware}.a) which we actively provide feedback on; to maximise brightness (i.e. the photon extraction rate), we control: (1) the single-photon source bias voltage, (2) the photon polarization at the input of the optical demultiplexer,  and (3) the photon polarization at each of the six inputs of the Ascella QPU; brightness is furthermore stabilized by (4) maintaining the excitation laser power at a fixed, pre-determined value; finally, to maintain photon arrival-time synchronicity (which is strongly related to two-qubit gate fidelity), (5) the fiber delays at the output of the optical demultiplexer are kept at constant temperature in a thermalized box. Each anomaly exceeding the scope of this stabilisation is used to develop more advanced simulation modules for each hardware component of the system, thereby increasing the automated test coverage.

Any set-up improvements and upgrades to the QPU hardware or software stack are then reviewed and discussed on a weekly basis, and implemented during the weekly 4-hour maintenance period.

While photonic integrated chips guarantee unprecedented compactness and stability for optical setups, they are subject to imperfections: passive phases \cite{Yang2015}, phase shifter thermal crosstalk \cite{Milanizadeh2019}, asymmetric directional couplers \cite{Bandyopadhyay2021} and inhomogeneous input/output losses \cite{Li2017}. We developed a machine learning-based characterization method that is able to measure these imperfections, which we can then effectively compensate. Our custom compilation process optimizes the applied on-chip phase shifts to enhance the fidelity of the implemented operation acting on the qubits. Subsequently, our tailored transpilation process translates the phase shifts into voltages used to control the photonic chip (see Fig. \ref{fig:main_hardware}.a), taking crosstalk in particular into account. The characterization process can take up to 24 hours for the \textit{Ascella} UPIC and must be executed every time a new single-photon source is installed, but it is also compatible with on-the fly updates, which is being implemented, leveraging ``usage data'' acquired during its operation time.


\subsection{Continuous benchmark and output analysis}
\label{sec:benchmark}
In addition to monitoring physical system parameters as presented in Section~\ref{sec:monitoring}, we also introduced very early high-level benchmarks in order to catch unanticipated system failures, and monitor the evolution of global performance over long periods of time. These benchmarks are designed to non-trivially use the full capacity of the processor to avoid blind spots in the monitoring: we compute 1-, 2- and 3-qubit gate fidelity using the protocol described in \cite{maring2023general} through respectively 4, 36 and 340 independent measurements. To avoid delay of user tasks, these measurements are launched by subset of measurements ran every 2 hours using period with no user activity. The resulting fidelity therefore aggregates measurements over long periods, incidentally highlighting any time-dependent fluctuations. Figure \ref{fig:main_hardware}.c shows for instance the monitoring of CNOT~\cite{ralph2002linear} through 12 full measurements per day over 120 days.

Finally, we exploit ``usage data'' by comparing processor output with simulators\footnote{The possibility to fully simulate output of QPUs is currently an opportunity for system understanding and development that we have to leverage fully since it will disappear when number of qubits will grow: for such photonic quantum computer, we expect to reach this point where output will not be anymore simulable for about 20 photons.} also available on the platform and using this comparison to progressively update noise models of the simulator and again to detect new types of system failures.



\subsection{User communication}

Up-to-date information regarding the hardware's performance is directly available on Quandela Cloud: the current status (available, in deployment, in calibration, in maintenance, or unreachable) of each computing platform, the number of tasks currently queuing, a detailed description of the hardware are available online, and the latest characterization of the key performance index. The performance metrics include the indistinguishability of the generated photons (Hong–Ou–Mandel visibility), the total transmittance of the setup and the single-photon purity ($\mathcal{P}=1-g^{(2)}(0)$). The physical constraints imposed by the hardware are also provided to the user. All this information can be used by the end-user to analyze outcome of his own algorithm. For example, the Hong–Ou–Mandel measurement is expected to have a direct impact on CNOT fidelity \cite{gazzano2013}. For each request, the user has therefore the ability to use this imperfection information for getting a better interpretation of his results.

Scheduled and unplanned maintenance periods are communicated through a \texttt{News} tab.

Finally, a support link is provided in the Cloud interface and dedicated personnel respond to any incoming request regardless of the user subscription plan.

\section{Conclusion}



This 6-month-long operational experience has been highly beneficial and instructive: it has allowed us to identify numerous points of failure and implement automated procedures, significantly reducing the need for ``human in the loop'' for troubleshooting and manual calibration. We have collected stability and performance benchmarks throughout the entire period, which has been invaluable for training better models and predicting complex failures or system erosion. Throughout this process, we have maintained close contact with end-users through direct and indirect communication.

Reaching the milestone of achieving \textit{first nine} availability is important to us, although it may be considered modest in comparison to HPC standards. Nonetheless, we believe that it should become a minimal requirement. Reaching the \textit{second nine} remains out of reach for the moment, given the inherent complexity of the hardware and the challenges in applying redundancy, which is a key aspect of achieving the highest availability in classical architectures.

In addition to operational improvements, we have also made significant progress in developing a data center-compatible solution adhering to 19-inch rack standards, as depicted in Figure \ref{fig:main_hardware}.b. This integrated system incorporates an alignment-free and mechanically vibration-insensitive pigtailed single-photon source, along with single photon detectors, within a compact and industrial \qty{4}{\kelvin} cryostat. The system architecture primarily relies on in-house opto-electronic modules, which utilizes a unified and centralized communication protocol and power management system. This design facilitates easy remote maintenance, incorporates auto-alignment and optimization functions, and eliminates the need for specialized operator intervention. Monitoring is an integral part of the system's design rather than an additional layer. The system has been intentionally built to incorporate monitoring capabilities from the ground up, ensuring efficient performance and reliability. Its compact and modular integration significantly mitigates environmental instabilities, allows for easy parts replacement, and serves as the foundation for future scaled-up versions of quantum computers.

While the journey towards full integration of Quantum Computers in full HPC environment and achieving quantum advantage in real-world applications just started and is expected to open new challenges, we advocate for a pragmatic incremental bottom-up approach. By transparently sharing each step with end-users, we can collectively acquire the necessary knowledge and methodologies for operating reliable quantum processors. High availability of the quantum processing unit (QPU) is crucial for successful integration into HPC environments but is not sufficient. Simultaneously, this integration process should be strongly guided by application-specific benchmarks to evaluate performance, optimize algorithms, and drive further advancements in quantum computing tailored to specific use cases.

\bibliographystyle{IEEEtran}
\bibliography{biblio}

\end{document}